\documentclass[journal]{IEEEtran}

\usepackage[usenames]{color}
\usepackage{amssymb}
\usepackage{graphicx}
\usepackage{amscd}

\usepackage{graphics,amsmath,amssymb,relsize}
\usepackage{bm}
\usepackage{amsthm}
\usepackage{amsfonts}
\usepackage{latexsym}
\usepackage{epsf}

\newtheorem{theorem}{Theorem}[section]
\newtheorem{corollary}[theorem]{Corollary}
\newtheorem{lemma}[theorem]{Lemma}
\newtheorem{proposition}[theorem]{Proposition}

\newtheorem{defin}[theorem]{Definition}
\newenvironment{definition}{\begin{defin}\normalfont\quad}{\end{defin}}
\newtheorem{examp}[theorem]{Example}

\newtheorem{rema}[theorem]{Remark}
\newtheorem{prob}[theorem]{Problem}

\numberwithin{equation}{section}

\newcommand{\bt}{\begin{thm}}
\newcommand{\et}{\end{thm}}
\newcommand{\bp}{\begin{proof}}
\newcommand{\ep}{\end{proof}}
\newcommand{\bprop}{\begin{prop}}
\newcommand{\eprop}{\end{prop}}
\newcommand{\bl}{\begin{lemma}}
\newcommand{\el}{\end{lemma}}
\newcommand{\bc}{\begin{corollary}}
\newcommand{\ec}{\end{corollary}}
\newcommand{\Z}{\mathbb{Z}}
\newcommand{\C}{\mathbb{C}}
\newcommand{\be}{\begin{enumerate}}
\newcommand{\ee}{\end{enumerate}}

\newcommand{\OMIT}[1]{}

\ifCLASSINFOpdf
  
\else
  
\fi

\begin{document}

\title{The Cayley graphs associated with some \\ quasi-perfect Lee codes are Ramanujan graphs}

\author{Khodakhast~Bibak, Bruce~M.~Kapron, and~Venkatesh~Srinivasan
\thanks{The authors are with the Department of Computer Science, University of Victoria, Victoria, BC, Canada V8W 3P6. Email: \{kbibak,bmkapron,srinivas\}@uvic.ca. Copyright (c) 2014 IEEE. Personal use of this material is permitted.  However, permission to use this material for any other purposes must be obtained from the IEEE by sending a request to pubs-permissions@ieee.org.}}

\maketitle

\begin{abstract}
Let $\Z_n[i]$ be the ring of Gaussian integers modulo a positive integer $n$. Very recently, Camarero and Mart\'{i}nez [IEEE Trans. Inform. Theory, {\bf 62} (2016), 1183--1192], showed that for every prime number $p>5$ such that $p\equiv \pm 5 \pmod{12}$, the Cayley graph $\mathcal{G}_p=\textnormal{Cay}(\Z_p[i], S_2)$, where $S_2$ is the set of units of $\Z_p[i]$, induces a 2-quasi-perfect Lee code over $\Z_p^m$, where $m=2\lfloor \frac{p}{4}\rfloor$. They also conjectured that $\mathcal{G}_p$ is a Ramanujan graph for every prime $p$ such that $p\equiv 3 \pmod{4}$. In this paper, we solve this conjecture. Our main tools are Deligne's bound from 1977 for estimating a particular kind of trigonometric sum and a result of Lov\'{a}sz from 1975 (or of Babai from 1979) which gives the eigenvalues of Cayley graphs of finite Abelian groups. Our proof techniques may motivate more work in the interactions between spectral graph theory, character theory, and coding theory, and may provide new ideas towards the famous Golomb--Welch conjecture  on the existence of perfect Lee codes.
\end{abstract}

\IEEEpeerreviewmaketitle

\section{Introduction}\label{Sec_1}

The long-standing Golomb--Welch conjecture \cite{GW} states that there are no perfect Lee codes for spheres of radius greater than 1 and dimension greater than 2. Resolving this conjecture has been one of the main motivations for studying perfect and quasi-perfect Lee codes. Very recently, Camarero and Mart\'{i}nez \cite{CM}, showed that for every prime number $p>5$ such that $p\equiv \pm 5 \pmod{12}$, the Cayley graph $\mathcal{G}_p=\textnormal{Cay}(\Z_p[i], S_2)$, where $S_2$ is the set of units of $\Z_p[i]$, induces a 2-quasi-perfect Lee code over $\Z_p^m$, where $m=2\lfloor \frac{p}{4}\rfloor$. They also conjectured \cite[Conj. 31]{CM} that the Cayley graph $\mathcal{G}_p=\textnormal{Cay}(\Z_p[i], S_2)$ is a Ramanujan graph for every prime $p$ such that $p\equiv 3 \pmod{4}$. In this paper, we solve this conjecture. Our main tools, which are reviewed in the next section, are Deligne's bound \cite{DEL} from 1977 for estimating a particular kind of trigonometric sum and a result of Lov\'{a}sz \cite{LO} from 1975 (or of Babai \cite{BA} from 1979) which gives the eigenvalues of Cayley graphs of finite Abelian groups. Our proof techniques may motivate more work in the interactions between spectral graph theory, character theory, and coding theory, and may provide new ideas towards the Golomb--Welch conjecture.

Let us first recall here briefly some terminologies and concepts that we will need in this paper. The ring of \textit{Gaussian integers} is defined as
$$
\Z[i]=\lbrace x+yi : x, y \in \Z, i=\sqrt{-1} \rbrace.
$$ 
In other words, Gaussian integers are the lattice points in the Euclidean plane. The \textit{norm} of a Gaussian integer $w=x+yi$ is N$(w)=|w|^2=x^2+y^2$. The elements of $\Z[i]$ with norm 1 are called the \textit{units} of $\Z[i]$; so, the units of $\Z[i]$ are just $\pm 1$ and $\pm i$. Similarly, the ring of Gaussian integers modulo a positive integer $n$ is defined as 
$$
\Z[i]/n\Z[i]\cong\Z_n[i]=\lbrace a+bi : a, b \in \Z_n, i=\sqrt{-1} \rbrace.
$$
Note that the definition of norm (and so unit) in the ring $\Z_n[i]$ is the same as that of $\Z[i]$ except that we need to evaluate the norm modulo $n$. That is, the \textit{norm} of $z=a+bi \in \Z_n[i]$ is N$(z)=a^2+b^2 \pmod{n}$, and $z=a+bi \in \Z_n[i]$ is a \textit{unit} of $\Z_n[i]$ if and only if 
$$
a^2+b^2 \equiv 1 \pmod{n}.
$$

The following classical result gives necessary and sufficient conditions under which the ring $\Z_n[i]$ is a field; see, e.g., \cite[Fact 3]{DRD}.

\begin{proposition}\label{field}
Let $n>1$ be an integer. The ring $\Z_n[i]$ is a field if and only if $n$ is a prime and $n\equiv 3 \pmod{4}$.
\end{proposition}

Let $\Gamma$ be a group written in additive notation. A non-empty subset $S \subseteq \Gamma$ is said to be \textit{symmetric} if $S=-S$, where $-S=\lbrace -x : x\in S \rbrace$. In other words, $S$ is symmetric if $-x \in S$ whenever $x \in S$. Now, we define Cayley graphs:

\begin{definition}
Let $\Gamma$ be a group, written additively, and $S$ be a finite symmetric subset of $\Gamma$ which does not contain the identity element of $\Gamma$. The \textit{Cayley graph} of $\Gamma$ with respect to $S$, denoted by $G=\text{Cay}(\Gamma, S)$, is the graph whose vertex set is $\Gamma$, and such that $u \sim v$ if and only if $v-u \in S$. Note that the Cayley graph $G=\text{Cay}(\Gamma, S)$ is undirected, simple, $|S|$-regular, and vertex-transitive. Also, $G$ is connected if and only if $S$ generates $\Gamma$.
\end{definition}

Roughly speaking, an expander is a highly connected sparse graph, that is, every subset of its vertices has a large set of neighbours. An important special case, namely, Ramanujan graphs are also of great interest. These graphs are actually `optimal' expanders, from the spectral point of view. Roughly speaking, a Ramanujan graph is a connected regular graph whose second largest eigenvalue in absolute value is `asymptotically' the smallest possible (or, equivalently, whose spectral gap is `asymptotically' the largest possible). Formally, a finite, connected, $k$-regular graph $G$ is called a \textit{Ramanujan graph} if every eigenvalue $\lambda\not=\pm k$ of $G$ satisfies the bound
$$
|\lambda|\leq 2\sqrt{k-1}.
$$

To this date, there are only a few explicit constructions (which are useful for applications) of expanders and Ramanujan graphs, all given using several strong (and seemingly unrelated!) mathematical tools; mainly from number theory. These graphs have a great deal of seminal applications in many disciplines such as computer science, cryptography, coding theory, and even in pure mathematics! See \cite{DSV, HLW, LUBO} for 
detailed discussions and surveys on expanders and Ramanujan graphs, their interactions with other areas like number theory and group theory, and their many wide-ranging applications.

Now, we review some basic facts about group characters; see, e.g., \cite{ISA, SE} for more details. A \textit{character} of a group $\Gamma$ is a group homomorphism from $\Gamma$ to the unit circle $S^1=\lbrace z \in \C : |z|=1 \rbrace$. So, if $\Gamma$ is a finite group then a character of $\Gamma$ can be defined as a group homomorphism from $\Gamma$ to $\C^{*}$, the multiplicative group of non-zero complex numbers. For a group $\Gamma$, the \textit{trivial} character $\chi_0$ is the function on $\Gamma$ where $\chi_0(g)=1$, for all $g\in \Gamma$. The characters of a finite group are linearly independent. A finite group $\Gamma$ has \textit{at most} $|\Gamma|$ characters and a finite Abelian group $\Gamma$ has \textit{exactly} $|\Gamma|$ distinct characters. For a finite Abelian group $\Gamma$ with the trivial character $\chi_0$,
\begin{equation*}
\sum_{g \in \Gamma}\chi(g) = 
\begin{cases}
|\Gamma|, & \text{ if \ $\chi=\chi_0$}; \\
0, & \text{ if \ $\chi\not=\chi_0$}.
\end{cases}
\end{equation*}

\section{Proof ingredients and techniques}\label{Sec_2}

In this section, we prove the conjecture proposed in \cite{CM}, by showing that the Cayley graph $\mathcal{G}_p=\textnormal{Cay}(\Z_p[i], S_2)$ is a $(p+1)$-regular Ramanujan graph. First, we mention the proof ingredients. The following proposition lists some classical facts from spectral graph theory; see, e.g., \cite{doob}. As it is common, by an eigenvalue (resp., eigenvector) of a graph we mean an eigenvalue (resp., eigenvector) of the adjacency matrix of that graph.

\begin{proposition}\label{facts} Let $G$ be a simple graph (i.e., without loops or multiple edges) of order $n$, with the adjacency matrix $A(G)$, and with the maximum degree $\Delta(G)$. Also, let $\lambda_{\min}(G)$ and $\lambda_{\max}(G)$ denote, respectively, the smallest and the largest eigenvalues of $G$. The following facts hold:

\begin{itemize}

\item The graph $G$ has $n$ eigenvalues (including multiplicities), and since $A(G)$ is real and symmetric, all these eigenvalues are real.

\item We have $\lambda_{\max}(G)\leq \Delta(G)$. Furthermore, if $G$ is $k$-regular then 
$\lambda_{\max}(G)=k$, and for every eigenvalue $\lambda$ of $G$, $|\lambda|\leq k$.

\item If $G$ is $k$-regular then the multiplicity of the eigenvalue $k$ equals the number of connected components of $G$. So, if $G$ is $k$-regular then $G$ is connected if and only if the eigenvalue $k$ has multiplicity one.

\item The graph $G$ is bipartite if and only if its spectrum is symmetric about \textnormal{0}. Also, if $G$ is connected then $G$ is bipartite if and only if $\lambda_{\min}(G)=-\lambda_{\max}(G)$.

\end{itemize}

\end{proposition}

It is well-known that the spectra of Cayley graphs of finite groups can be expressed in terms of characters of the underlying group (\cite{BA, LO}). The following result determines the eigenvalues and
eigenvectors of Cayley graphs of finite Abelian groups. The theorem follows from a more general result of Lov\'{a}sz \cite{LO} from 1975 (or of Babai \cite{BA} from 1979).

\begin{theorem}\label{eigenvalue}
Let $\Gamma$ be a finite Abelian group, $\chi : \Gamma \rightarrow \C^{*}$ be a character of $\Gamma$, and $S$ be a symmetric subset of $\Gamma$ which does not contain the identity element of $\Gamma$. Then the vector $v_{\chi}=(\chi(g))_{g \in \Gamma}$ is an eigenvector of the Cayley graph $G=\textnormal{Cay}(\Gamma, S)$, with the corresponding eigenvalue being 
$$
\lambda_{\chi}=\sum_{s \in S}\chi(s).
$$
\end{theorem}

In order to find the degree of the Cayley graph $\mathcal{G}_p=\textnormal{Cay}(\Z_p[i], S_2)$, we need to evaluate the number of solutions of certain quadratic congruences. The problem of counting the number of solutions of quadratic congruences in several variables has been investigated, in a general form, in \cite{TOTH}, where a general formula is proved. Specifically, T\'oth \cite{TOTH} considered the quadratic congruence
\begin{equation} \label{quadratic cong}
a_1x_1^2+ \cdots +a_kx_k^2 \equiv b \pmod{n},
\end{equation}
where $b\in \Z$, $\mathbf{a}=(a_1,\ldots,a_k)\in \Z^k$, and proved an explicit formula (see Theorem~\ref{main quadratic} below) for the number $N_k(b,n,\mathbf{a})$ of solutions $\langle x_1,\ldots,x_k\rangle\in \Z_n^k$ of \eqref{quadratic cong}, when $n$ is odd. The formula involves a special kind of trigonometric sums, namely, quadratic Gauss sums that we now define. Let $e(x)=\exp(2\pi ix)$ be the complex exponential with period 1. For positive integers $m$ and $n$ with $\gcd (m,n)=1$, the quantity 
\begin{align}\label{Gauss Sum}
S(m,n)= \mathlarger{\sum}_{j=1}^n e\left(\frac{m j^2}{n}\right)
\end{align}
is called a {\it quadratic Gauss sum}.

\begin{theorem} \label{main quadratic} Let $k$, $b$, $n$ be integers \textnormal{(}$k,n \geq 1$\textnormal{)}, and $\mathbf{a}=(a_1,\ldots,a_k)\in \Z^k$. We have
\begin{align*}
\ & N_k(b,n,\mathbf{a})\\
=\ & n^{k-1} \mathlarger{\sum}_{d\;\mid\;n} \frac1{d^k} \mathlarger{\sum}_{\substack{m=1\\
(m,d)=1}}^d e\left(\frac{-bm}{d}\right) S(m a_1,d) \cdots S(m a_k,d).
\end{align*}
\end{theorem}

Putting $k=2$, $a_1=a_2=1$, $b=1$, and $n=p^r$ (a power of a prime) in Theorem~\ref{main quadratic}, the following special case is obtained (see \cite{TOTH}):

\begin{lemma}\label{quadratic} Let $p$ be a prime and $r$ be a positive integer. The number $N_2(1,p^r)$ of solutions of the quadratic congruence $x^2+y^2\equiv 1 \pmod{p^r}$ is
\begin{equation*}
N_2(1,p^r) = \begin{cases}
p^r (1-\frac{1}{p}), & \text{if \ $p \equiv 1$ {\rm (mod $4$)}, $r \ge 1$}; \\
p^r (1+\frac{1}{p}), & \text{if \ $p \equiv 3$ {\rm (mod $4$)}, $r \ge 1$}; \\
2, & \text{if \ $p=2$, $r=1$}; \\
2^{r+1}, & \text{if \ $p=2$, $r \ge 2$}.
\end{cases}
\end{equation*}
\end{lemma}

If $\mathbb{F}$ and $\mathbb{E}$ are fields and $\mathbb{F} \subseteq \mathbb{E}$, then $\mathbb{E}$ is said to be an {\it extension} of $\mathbb{F}$, denoted by $\mathbb{E}\;/\;\mathbb{F}$. The {\it degree} of a field extension $\mathbb{E}\;/\;\mathbb{F}$, denoted by $[\mathbb{E} \;:\; \mathbb{F}]$, is the dimension of $\mathbb{E}$ as a vector space over $\mathbb{F}$. A field extension $\mathbb{E}\;/\;\mathbb{F}$ is called a \textit{finite extension} if $[\mathbb{E} \;:\; \mathbb{F}]<\infty$. Let $\mathbb{F}_{q^n}$ be a finite extension field of the finite field $\mathbb{F}_q$. For $\alpha \in \mathbb{F}_{q^n}$, the \textit{field norm} of $\alpha$ is defined by (see, e.g., \cite[Def. 2.27]{LIN})
$$
\text{N}_{\mathbb{F}_{q^n}/\mathbb{F}_q}(\alpha)=\alpha^{(q^n-1)/(q-1)}.
$$
The elements of $\mathbb{F}_{q^n}$ with field norm 1 are called the \textit{units} of $\mathbb{F}_{q^n}$. 

\begin{lemma}\label{norm}
Let $p$ be a prime such that $p\equiv 3 \pmod{4}$. Then for every $z\in \Z_p[i]$ the field norm of $z$ coincides with the norm of $z$ in the usual sense, that is, as the norm of a Gaussian integer modulo $p$.
\end{lemma}

\begin{proof}
Let $z=a+bi \in \Z_p[i]$, where $p$ is a prime and $p\equiv 3 \pmod{4}$. Then, by the above definition, the field norm of $z$ equals 
\begin{eqnarray*}
  \text{N}_{\Z_p[i]/\Z_p}(a+bi)&=&(a+bi)^{p+1}\\
  &=&(a+bi)(a+bi)^{p}\\
  &=&(a+bi)(a^p+b^pi^p)\\
  &=&a^{p+1}+ab^pi^p+ba^pi+b^{p+1}i^{p+1}\\
  &\equiv &a^2+b^2 \pmod{p},
\end{eqnarray*}
where we have used Fermat's little theorem and also the binomial theorem for commutative rings of characteristic $p$ (see, e.g., \cite[Th. 1.46]{LIN}) which says that in a commutative ring $R$ of prime characteristic $p$, we have 
$$
(x+y)^{p^n}=x^{p^n}+y^{p^n},
$$ 
for every $x,y\in R$ and every positive integer $n$. Note that the value $a^2+b^2 \pmod{p}$ is just the norm of $z$ as a Gaussian integer modulo $p$.
\end{proof}

Deligne \cite{DEL} using tools from algebraic geometry and cohomology proved the following crucial bound.

\begin{theorem}\label{Deligne's bound}
Suppose that $\mathbb{F}_{q^n}/\mathbb{F}_q$ is the field extension of degree $n$ of the finite field $\mathbb{F}_q$, $S_n$ is the set of units of $\mathbb{F}_{q^n}$, and $\chi$ is a nontrivial character of the additive group of $\mathbb{F}_{q^n}$. Then
$$
\Big|\sum_{s \in S_n}\chi(s)\Big|\leq nq^{\frac{n-1}{2}}.
$$
\end{theorem}

Now, we are ready to prove our main result. This problem has been mentioned as Conjecture 31 in \cite{CM}.

\begin{theorem}\label{main}
Let $p$ be a prime, $p\equiv 3 \pmod{4}$, and $S_2$ be the set of units of $\Z_p[i]$. Then the Cayley graph $\mathcal{G}_p=\textnormal{Cay}(\Z_p[i], S_2)$ is a $(p+1)$-regular Ramanujan graph.
\end{theorem}

\begin{proof}
By Proposition~\ref{field}, the ring $\Z_n[i]$ is a field if and only if $n$ is a prime and $n\equiv 3 \pmod{4}$. Thus, for a prime $p$ with $p\equiv 3 \pmod{4}$ we have $\Z_p[i]\cong \mathbb{F}_{p^2}$. Also, we know that for a prime $p$ with $p\equiv 3 \pmod{4}$, $\Z_p[i]$ as an extension field of the finite field $\mathbb{F}_p$ has degree 2 (because $\lbrace 1, i\rbrace$ can serve as a basis), that is, $[\Z_p[i]:\mathbb{F}_p]=2$.

Note that $S_2$ is a symmetric subset of $\Z_p[i]$ and does not contain the identity element of $\Z_p[i]$. Since the Cayley graph $\mathcal{G}_p=\text{Cay}(\Z_p[i], S_2)$ is of order $p^2$, it has $p^2$ real eigenvalues. Also, by Lemma~\ref{quadratic}, the number of solutions of the quadratic congruence $x^2+y^2\equiv 1 \pmod{p}$ is $p+1$, so, $|S_2|=p+1$ which means that $\mathcal{G}_p$ is $(p+1)$-regular. By Theorem~\ref{eigenvalue}, the eigenvalues of $\mathcal{G}_p$ are determined by 
$$
\lambda_{\chi}=\sum_{s \in S_2}\chi(s),
$$
where $\chi$ runs over all characters of $\Z_p[i]$; note that since $\Z_p[i]$, as an additive group, is a finite Abelian group, it has $p^2$ distinct characters. The eigenvalue corresponding to the trivial character $\chi_0$ of $\Z_p[i]$ equals   
$$
\lambda_{\chi_0}=\sum_{s \in S_2}\chi_0(s)=\sum_{s \in S_2}1=|S_2|=p+1.
$$
Of course, as $\mathcal{G}_p$ is $(p+1)$-regular, we already knew, by Proposition~\ref{facts}, that $p+1$ is an eigenvalue of $\mathcal{G}_p$ (in fact, the largest one). 

Note that since $p$ is a prime and $p\equiv 3 \pmod{4}$, by Lemma~\ref{norm}, for every $z\in \Z_p[i]$ the field norm of $z$ coincides with the norm of $z$ as a Gaussian integer modulo $p$, thus, the `field norm' (and so unit) in Theorem~\ref{Deligne's bound} is in fact the `norm' (and so unit) we already have. Now, by Theorem~\ref{Deligne's bound}, the absolute values of the eigenvalues corresponding to the nontrivial characters $\chi\not=\chi_0$ of $\Z_p[i]$ satisfy the bound  
$$
|\lambda_{\chi}|=\Big|\sum_{s \in S_2}\chi(s)\Big|\leq 2\sqrt{p}.
$$
Therefore, $\mathcal{G}_p$ is a $(p+1)$-regular Ramanujan graph. We remark that since $\mathcal{G}_p$ is $(p+1)$-regular and the eigenvalue $p+1$ has multiplicity one, by Proposition~\ref{facts}, $\mathcal{G}_p$ is connected. This in turn implies that $S_2$ generates $\Z_p[i]$.
\end{proof}

Since by the above argument, $-(p+1)$ is not an eigenvalue of $\mathcal{G}_p$, by Proposition~\ref{facts}, we get:

\begin{corollary}
The Cayley graph $\mathcal{G}_p=\textnormal{Cay}(\Z_p[i], S_2)$ is not bipartite. This implies that
$\mathcal{G}_p$ has at least one odd cycle.
\end{corollary}

\section*{Acknowledgements}

During the preparation of this work the first author was supported by a Fellowship from the University of Victoria (UVic Fellowship).

\begin{IEEEbiographynophoto}{Khodakhast Bibak}
is a PhD student at the Department of Computer Science, University of Victoria. He obtained an MMath degree at the Department of Combinatorics \& Optimization, University of Waterloo in 2013. His research interests are mainly cryptography, information security, information theory, discrete mathematics, and number theory.
\end{IEEEbiographynophoto}

\begin{IEEEbiographynophoto}{Bruce Kapron}
is a Professor in the Computer Science Department at the University of Victoria. He received a B.Math. in Computer Science and Pure Mathematics from the University of Waterloo in 1984, a M.Sc. in Mathematics from Simon Fraser University in 1986 and a Ph.D. in Computer Science from the University of Toronto in 1991. His research interests include security, foundations of cryptography, logic, verification, and computational complexity.
\end{IEEEbiographynophoto}

\begin{IEEEbiographynophoto}{Venkatesh Srinivasan}
received a B.Eng degree from Birla Institute of Technology and Science, India, in 1994, and a Ph.D. degree from Tata Institute of Fundamental Research, India, in 2000. He joined the Department of Computer Science at the University of Victoria, Canada, in 2003 and is currently an Associate Professor there.  His research interests include algorithms for data analytics and data privacy, computational complexity and its connections with cryptography.
\end{IEEEbiographynophoto}

\end{document}